\documentclass[%
 reprint,
 amsmath,amssymb,
 aps,
]{revtex4-1} 

\usepackage{graphicx}
\usepackage{dcolumn}
\usepackage{bm}
\usepackage{color}
\usepackage{amssymb}
\begin{document}
\newcommand{\bor}{borophene}
\newcommand{\borh}{borophane}
\newcommand{\mose}{MoSe$_2$}
\newcommand{\wse}{WSe$_2$}
\newcommand{\mos}{MoS$_2$}
\newcommand{\ws}{WS$_2$}

\title{Tuning the p-type Schottky barrier in 2D metal/semiconductor 
interface: boron-sheet/\mose, and /\wse}

\author{W. R. M. Couto, and R. H. Miwa}
\email[]{hiroki@ufu.br}
\affiliation{Instituto de F\'isica, Universidade Federal de 
Uberl\^andia, CP 593, 38400-902, Uberl\^andia, MG, Brazil}
\affiliation{Instituto Federal do Tri\^angulo Mineiro, 
38600-000, Paracatu, MG, Brazil.}
\author{A. Fazzio}
\affiliation{Instituto de F\'isica,
Universidade de S\~ao Paulo, CP 66318, 05315-970, S\~ao Paulo, SP,
Brazil.}
\affiliation{Centro de Ci\^encias Naturais e Humanas, Universidade 
Federal do ABC, 09210-170, Santo Andr\'e, SP, Brazil.}

\date{\today}

\begin{abstract}

The electronic and the structural properties of two dimensional van der Waals 
metal/semiconductor heterostructures have been investigated through 
first-principles calculations. We have considered the  recently synthesized 
borophene  [Science {\bf 350}, 1513 (2015)], and the planar boron sheets (S1 and 
S2)  [Nature Chemistry {\bf 8}, 563 (2016)] as the 2D metal layer, and the 
transition metal dichalcogenides (TMDCs) \mose, and \wse\ as the semiconductor 
monolayer. We find that the energetic stability of those 2D metal/semiconductor 
heterojunctions is mostly ruled by the vdW interactions; however, chemical 
interactions also take place in borophene/TMDC. The electronic charge transfers 
at the metal/semiconductor interface has been mapped, where we find a a net 
charge transfer from the TMDCs to the boron sheets. Further electronic structure 
calculations reveal that the metal/semiconductor interfaces, composed by planar 
boron sheets S1 and S2, present a p-type Schottky barrier which can be tuned to 
a p-type ohmic contact upon an external electric field.
 
\end{abstract}

\pacs{.}
\keywords{...}

\maketitle

\section{Introduction}

Two dimensional (2D) systems have been the subject of numerous  studies 
addressing not only the fundamental research but also technological applications 
focusing the development of electronic devices. Such a 2D scenario in the 
material science was  predicted by Geim and Grigorieva\,\cite{geimNature2013}, 
where they present the concept of van der Waals (vdW) heterostructures, and  
more recently by Novoselov {\it et al.}\,\cite{novoselovScience2016}. Here, the 
electronic properties of those vdW heterostructures can be tuned  by  
stacking different sets of layered, semiconductor or metallic, materials.

Since the successful synthesis of single layer \mos\ 
transistor\,\cite{radisavljevicNatNanotech2011},  transition metal 
dichalcogenides (TMDCs) have been considered quite promising to be used in 2D 
vdW heterostructures. For instance, semiconductor/semiconductor vdW  
heterojunction composed by layered \mose/\wse\,\cite{huangNatMat2014}, and 
\mos/\ws\,\cite{gongNatMat2014}. Here a formidable bandgap engineering can be 
done through a suitable choice of the stacked  2D materials, as well as the 
number of the stacked layers. Based on first-principles calculations,  Kang {\it 
 et al.}\,\cite{kangAPL2013} performed an extensive study of the band offsets of 
TMDCs. The band offsets were estimated by comparing the ionization potentials 
and the band gaps of the stacked 2D components, showing that \mose/\wse, 
\mos/\ws\ and \mos/\wse\ heterojunctions present  type-II band alignment. 
Indeed, such a type-II band offset was verified  in a recent experimental 
realization of  \mos/\wse\ p-n heterojunction\,\cite{leeNatNanotech2014}. 
Meanwhile, metal/semiconductor 2D vdW heterostructures  have been successfully 
synthesized  through deposition of 2D (semi)metal on layered TMDC, for instance, 
graphene  on TMDCs\,\cite{yuNanoLett2014,shihACSNano2014}, or 2D metallic TMDCs, 
like H-NbS$_2$, on semiconductor TMDCs\,\cite{liuSciAdv2016}. 

In vdW heterostructures,  the Schottky barrier ($\Phi_{\rm B}$)  can be 
estimated by comparing the work function of the metal and the electronic 
affinity (n-type $\Phi_{\rm B}$) or the ionization potential (p-type $\Phi_{\rm 
B}$) of the semiconductor. Similarly to their 3D counterpart, the height of the 
Schottky barrier can be tuned by an external electric field, as observed in 
field effect transistors (FETs). However,  the hole injection in FETs based on 
(semiconductor) TMDC has been limited by the larger values of the p-type 
$\Phi_{\rm B}$. There are some of proposals aiming to provide an efficient hole 
injection in TMDCs; for instance, by using the oxidized graphene at the 
source/drain contacts\,\cite{chuangNanoLett2014mos2,mussoACSNano2014}, and more 
recently the inclusion of a BN monolayer at the metal/semiconductor 
interface\,\cite{farmanbarPRB2015}. However, the control of the   p-type 
$\Phi_{\rm B}$,  in order to  get an efficient hole injection in the metal/TMDCs 
heterojunctions, is still a challenge. Fortunately, nowadays we are facing an 
intense research on the  new materials, including 2D crystals, allowing to make 
a number of material combinations focusing on a given electronic 
property; for instance the p-type Schottky barrier in 
metal/semiconductor-TMDCs. 

Very recently metallic 2D boron sheets lying on the Ag(111) surface have been 
successfully synthesized, where one is characterized by vertically buckled  
boron atoms, hereafter named S0 
[Fig.\,\,\ref{pristine}(a1)]\,\cite{mannixScience2015}. While the others, upon 
the presence of boron vacancies, are flat [S1 and S2 in 
Figs.\,\ref{pristine}(b1) and (c1)]\,\cite{fengNatChem2016}. Here one can guess 
that those 2D boron sheets may act as the metal contact in 2D vdW 
heterostructures. In  a recent theoretical study\,\cite{camilo2DMat2017}, 
supported by the experimental findings\,\cite{mannixScience2015}, the authors 
verified that the  oxidation of S0 is quite likely, giving rise to boron 
vacancies, which somewhat mimic the planar geometries of S1 and S2,  as verified 
by Feng {\it et al.}\,\cite{fengNatChem2016}. Beside the inertness with respect 
to the oxidation process, it has been reported that the boron sheets S1 
and S2 are energetically more stable than the buckled geometry of borophene, S0.

In this work, based on the first-principles calculations, we investigate  the 
energetic stability and the electronic properties of 2D metal/semiconductor 
heterojunctions. We have considered the recently synthesized boron sheets 
S0--S2\,\cite{mannixScience2015,fengNatChem2016} as the 2D metallic layer; and 
monolayers (MLs) of TMDCs \mose\ and \wse\ as the semiconductor system. Our 
total energy results show that the energetic stability of those  
metal/semiconductor interfaces is mostly dictated by the vdW interactions. 
Further electronic band structure calculations reveal  the formation of 
interface (metallic) states in S0/TMDC,  pinning the Fermi level within the 
energy gap of the TMDC. On the other hand,  S1/ and S2/TMDC heterojunctions 
present a tunable p-type Schottky barrier; where we show that the former system 
presents a p-type ohmic contact upon an external electric field smaller than 
4\,V/nm.

\section{Method}

The calculations were performed using the density functional theory (DFT) as 
implemented in the VASP code\,\cite{vasp1,vasp2}. The exchange-correlation 
potential was described within the generalized gradient approximation 
(GGA-PBE)\,\cite{PBE}, and the electron-ion interactions were treated by using 
the the projected augmented wave approach (PAW)\,\cite{paw,kressePRB1999}. The 
Kohn-Sham wave functions were expanded in a plane-wave basis set with an energy 
cutoff of 400\,eV. The atomic positions were fully relaxed by including the van 
der Waals interactions by using the self-consistent opt88vdW 
approach\,\cite{dionPRL2004,klimesJPhysC2010}.  We have considered a force 
convergence tolerance of 20\,meV/\AA. In order to verify the validity of our 
results, the electronic structure of the pristine systems, {\it viz.}: single 
layer \mose, \wse,  and the boron sheets S0--S2 were calculated 
using HSE06 hybrid functional\,\cite{heydJChemPhys2006}. The 
metal/semiconductor interfaces were described within the slab method, where we 
have introduced a vacuum region of $\sim$15\,\AA, perpendicular to the 
metal/semiconductor interface,  in order to prevent spurious  interactions 
between a given interface and its (periodic) image. The Brillouin zone sampling 
was performed by using a set of 12$\times$12$\times$1 k-points within the 
Monkhorst-Pack scheme\,\cite{mp}.

\section{Results}

 \begin{figure}
  \includegraphics[width=8.6cm]{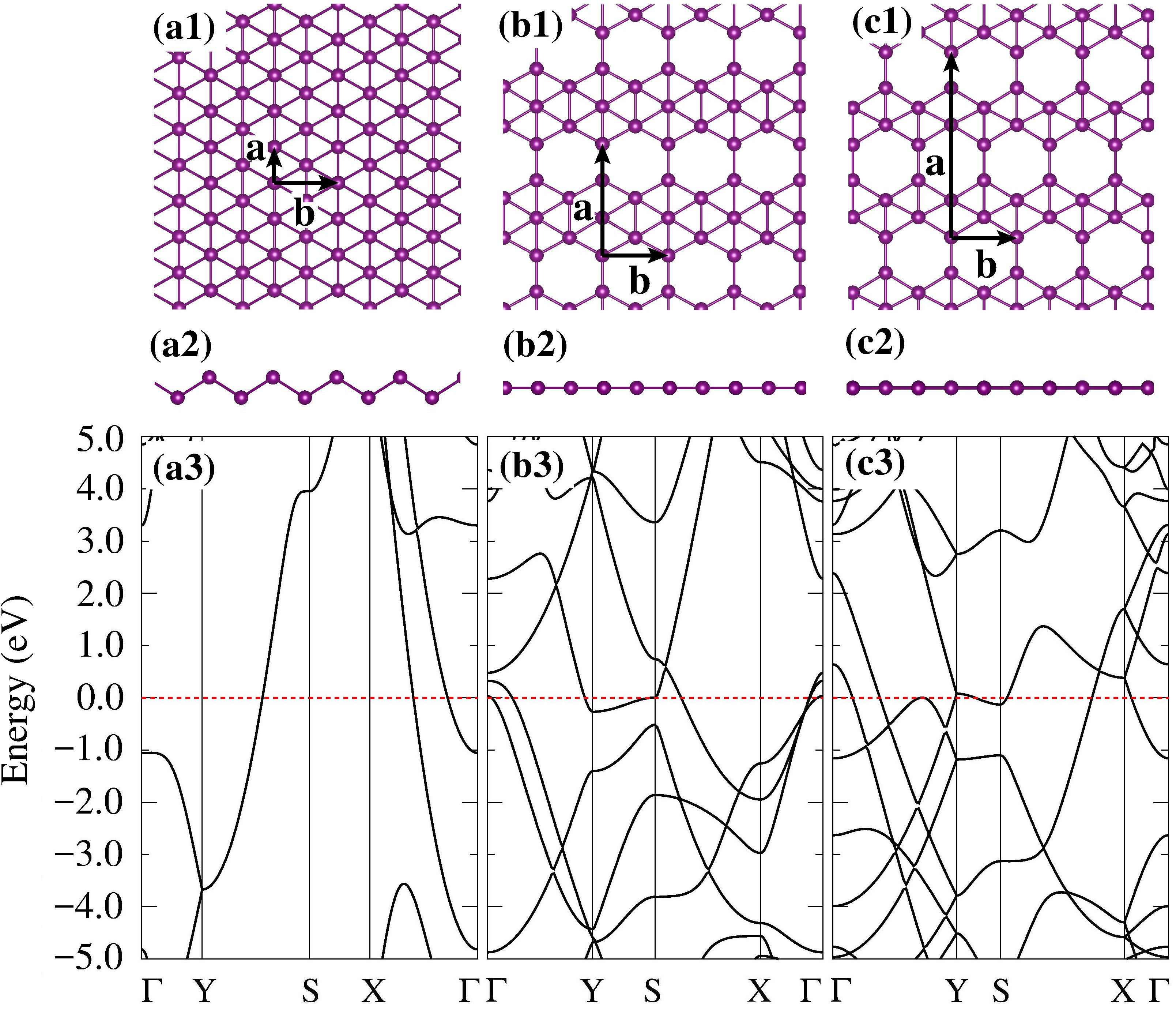}
  \caption{(Color online) Structural models and the electronic band structure 
of the pristine \bor\ (a), and the boron sheets S1 (b), and S2 (c). In 
(a3)--(c3) the Fermi level is set to zero.}
 \label{pristine}
 \end{figure}
 
In Fig.\,\ref{pristine}  we present the structural models and the electronic 
band structures of pristine free-standing boron sheets S0--S2, 
Figs.\,\ref{pristine}(a)--(c).  In agreement with the recent experimental 
findings, boron atoms in S0 [Fig.\,\ref{pristine}(a1)] present a vertical 
buckling of 0.91\,\AA, giving rise to boron stripes along the $\rm\bf\hat{a}$ 
direction. Meanwhile, upon the formation of boron vacancies, the structural 
models  S1 and S2 become planar, Fig.\,\ref{pristine}(b1) and (b2). The 
electronic band structure of S0, Fig.\,\ref{pristine}(a3), presents a set of 
metallic bands for wave vectors parallel to the $\rm\bf\hat{a}$ direction, that 
is along the boron stripes; whereas there are no metallic bands for wave vectors 
along the $\Gamma$Y abd SX directions. Such a band structure anisotropy is 
reduced in S1 and S2, Figs.\,\ref{pristine}(b3) and (c3). The energy bands of 
boron sheets S1 and S2 are characterized by  the formation of dispersionless 
metallic bands along the YS, and higher density of states (DOS) near the Fermi 
level ($\rm E_F$), when compared with the DOS of S0. In a previous work, we find 
that the anisotropy of the energy bands in \bor\ promotes a peculiar directional 
dependence of its electronic transport properties\,\cite{padilhaPCCP2016}. Here 
we will examine how those energy bands change upon the formation of 2D 
metal/semiconductor heterostructures, and their role on the Schottky barrier.

 \begin{figure}
  \includegraphics[width=8.6cm]{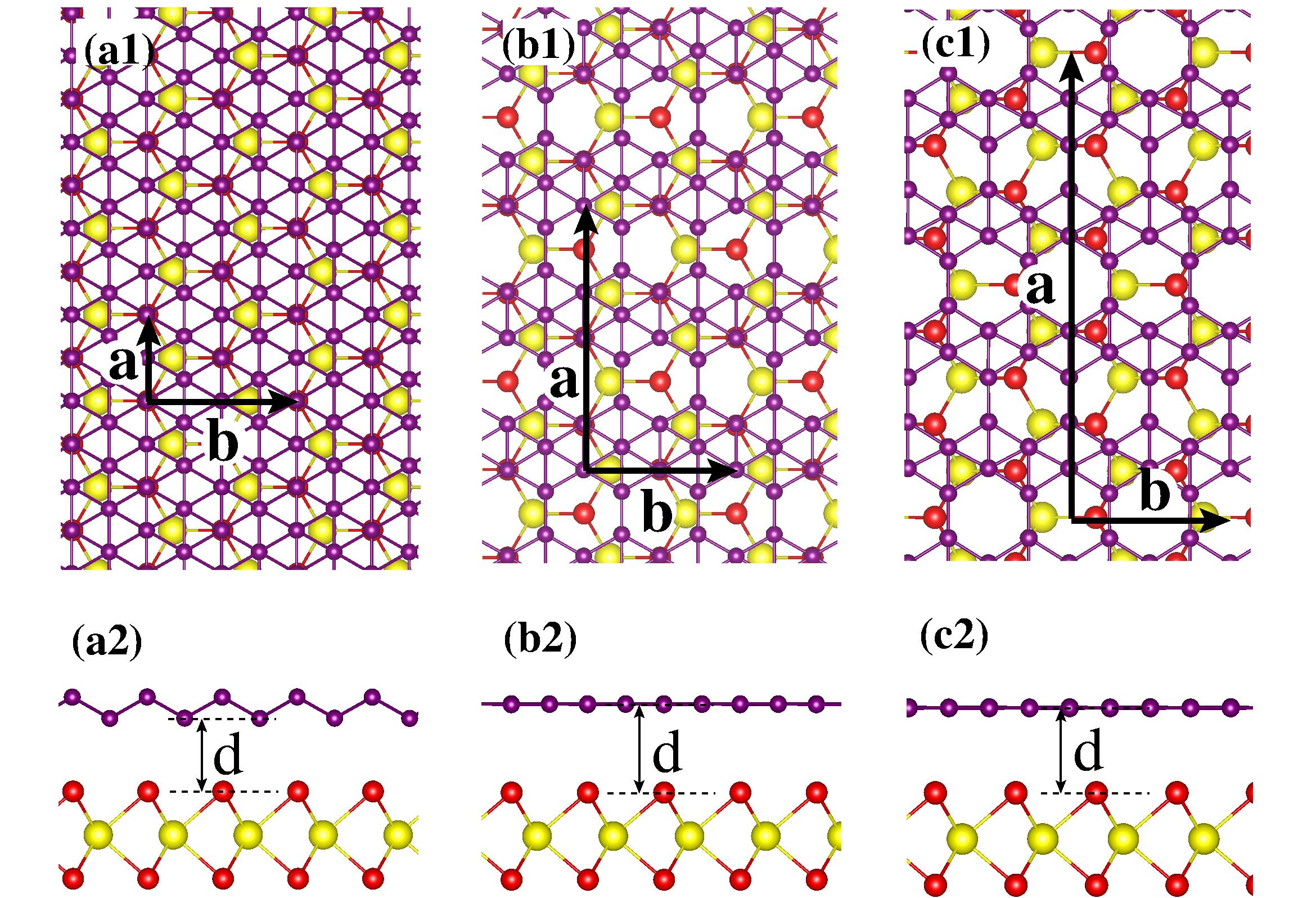}
  \caption{(Color online) Structural models of S0/TMDC top view (a1) and side 
view (a2), S1/TMDC top view (b1) and side view (b2),   S2/TMDC top view (c1) 
and side view (c2).}
 \label{hetero}
 \end{figure}

In  Fig.\,\ref{hetero} we present the structural models of metal/semiconductor  
systems (semiconductor = TMDCs \mose\ and \wse\ monolayers), for metal  = S0 
(a), S1 (b), and S2 (c). In order to minimize the lattice mismatch between the 
boron sheets and the TMDC monolayers, the former were described by orthorhombic 
supercells with surface periodicities of (2$\times$2). In this case, the boron 
sheets are strained by up to $\pm2.2$\%  with respect to their equilibrium 
lattice constant. For the  \mose\ and \wse\ monolayers (MLs) we have considered 
orthorhombic unit cells with the following surface periodicities, (1$\times$1), 
(3$\times$1), and (5$\times$1) for S0/\mose\ (S0/\wse), S1/\mose\ (S1/\wse), and 
S2/\mose\ (S2/\wse), respectively. In this case, the \mose\ and \wse\ monolayers 
are strained by up to $\pm0.9$\%, when compared with their equilibrium lattice 
constant.

The energetic stability of 2D heterostructures was inferred through the 
calculation of the metal/TMDC binding energy ($E^b$), defined as,
$$
E^b = E[{\rm metal}] + E[{\rm TMDC}] - E[{\rm metal/TMDC}],
$$
where $E[{\rm metal}]$ and $ E[{\rm TMDC}]$ represent the total energies of the 
separated systems, boron sheets and the TMDCs; and $E[{\rm metal/TMDC}]$ is the 
total energy of the final system, metal/TMDC 2D heterojunction.  We have 
considered four different metal/TMDC interface configurations, where we found 
$E^b$ between 34 and 44\,meV/\AA$^2$ for S0/\mose, and $E^b=29-41$\,meV/\AA$^2$ 
for S0/\wse. At the equilibrium geometry, the borophene sheet S0 and \mose\ 
(\wse) monolayer presents a vertical distance ($d$) between 2.79 (2.90) and 
3.17\,\AA\ (3.55\,\AA). Those findings of $E^b$  and $d$ are somewhat comparable 
with the ones obtained for bilayer-graphene on the Cu(111) 
surface\,\cite{eversonPRB2016}, namely $E^b$ of about 39\,meV/\AA$^2$, and 
vertical distance of  $\sim$2.9\,\AA; which allow us to infer that the energetic 
stability of borophene/TMDC systems is mostly dictated by vdW interactions. 
Indeed, by turning off the vdW contribution, we obtained $E^b$ of 
0.13\,meV/\AA$^2$ and $d$=3.97\,\AA. However, as will be discussed below,  the 
chemical interaction between S0 and the TMDC can not be neglected for 
$d<3$\,\AA.  The binding energy reduces for the planar boron sheets S1 and S2, 
namely we obtained $E^b$=33 and 31\,meV/\AA$^2$ for S1/ and S2/\mose\ 
respectively, and $d>3$\,\AA.  Our results of binding energies and equilibrium 
geometries,  for the energetically most stable configurations, are summarized in 
Table~I.

\begin{table}[h]
\caption{\label{energy} Results of  binding energy ($E^b$ in meV/\AA$^2$), and 
 equilibrium vertical distance ($d$ in \AA,  Fig.\,\ref{hetero}) of S0/TMDC, 
S1/TMDC, and S2/TMDC.}
\begin{ruledtabular}
\begin{tabular}{lcc}
metal/semic. &  $E^b$    &  $d$  \\     
\hline
  S0/\mose\  &  44   & 2.79   \\
  S0/\wse\   &  41   & 2.90  \\
 S1/\mose\ &  33   & 3.33  \\
 S1/\wse\  &  31   & 3.23  \\
 S2/\mose\ &  31   & 3.36  \\
 S2/\wse\  &  30   & 3.39  \\
\end{tabular}
\end{ruledtabular}
\end{table}

 \begin{figure}
  \includegraphics[width=8.6cm]{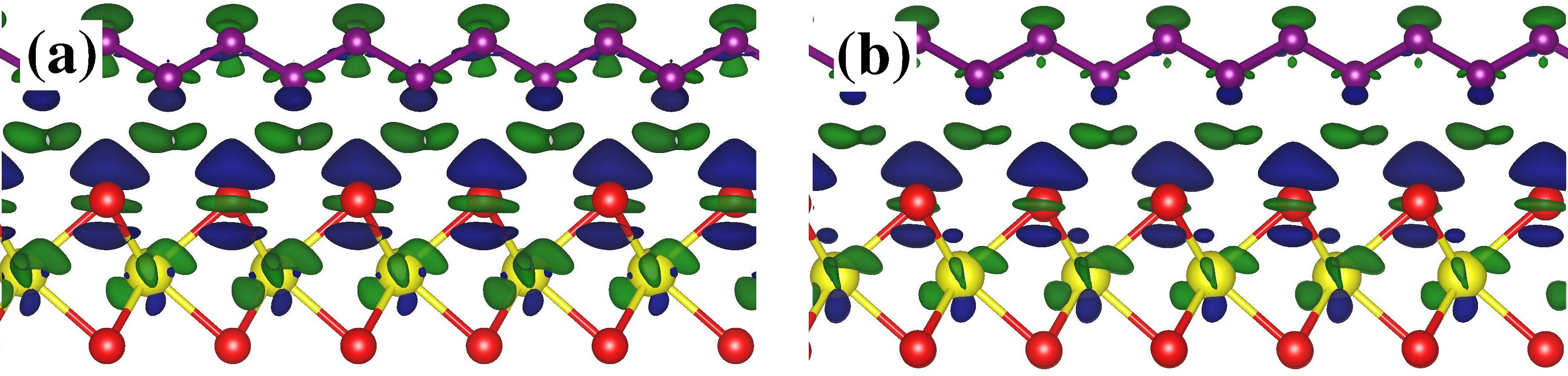}
  \caption{(Color online) Total charge transfers, $\Delta\rho$, at the S0/\mose\ 
(a), and S0/\wse\ (b) interface. Green regions indicate a net charge gain 
($\Delta\rho>0$), and blue regions indicate a net charge loss ($\Delta\rho<0$) 
with respect to the isolated components (Isosurfaces of $5\times10^{-4}$ 
e/\AA$^3$).}
 \label{drho-B}
 \end{figure}

Upon the formation of metal/TMDC heterostructures, there is a net 
charge transfer ($\Delta\rho$) at the metal/TMDC interface region. We 
may have a spacial picture of $\Delta\rho$ by comparing the total charge 
densities of the final system ($\rm\rho[metal/TMDC]$) with the ones of the 
(initial) separated components, metallic boron sheet ($\rm\rho[metal]$) and 
semiconductor \mose\ and \wse\ ($\rm\rho[TMDC]$), 
$$
\rm\Delta\rho = \rho[metal/TMDC]-\rho[metal]-\rho[TMDC].
$$ 

In Figs.\,\ref{drho-B}(a) and (b) we present our result of $\Delta\rho$ for 
S0/\mose\ and S0/\wse, where we verify that the interface region near the Se 
layer presents a  charge density loss ($\Delta\rho<0$), whereas near the 
borophene layer we find (mostly) $\Delta\rho>0$. Based on the Bader charge 
density analysis\,\cite{bader}, we find that the total charge density of the 
\mose\ and \wse\ monolayer reduces by 0.75 and $1.1\times10^{13}$\,e/cm$^2$. 
That is, indeed there is a net charge transfer from the boron sheet S0 to the 
\mose\ and \wse\ MLs.

\begin{figure}
  \includegraphics[width=8.6cm]{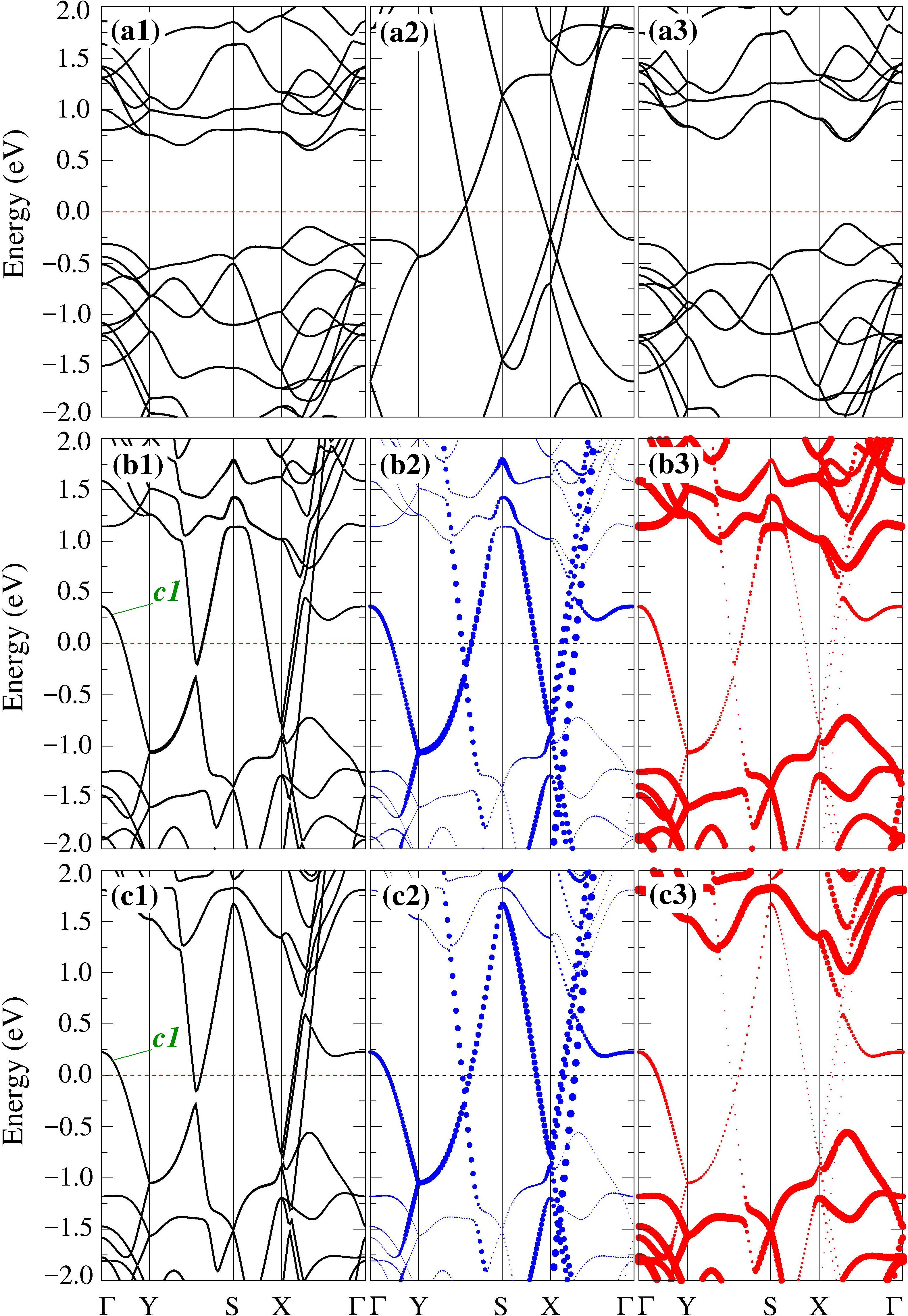}
  \caption{(Color online) Electronic band structures of pristine systems,  
(1$\times$1) \mose\ (a1),  \wse\ (c1), and (2$\times$2) borophene sheet S0 (b1). 
Electronic band structures of the S0/\mose\ (b1), projected on S0 (b2), and on 
the \mose\ ML (b3); S0/\wse\ (c1), projected on S0 (c2), and on the \wse\ ML 
(c3). The Fermi level is set to zero.}
 \label{banda-B_sem}
 \end{figure}

Figures\,\ref{banda-B_sem}(a1)--(a3) show the electronic band structures of the 
pristine systems considering the (1$\times$1) orthorhombic unit cells for \mose\ 
and \wse, and (2$\times$2) for the boron sheet S0, confirming the semiconductor 
(metallic)  character of \mose\ and \wse\ (S0). In Figs.\,\ref{banda-B_sem}(b1) 
and (c1) we present the electronic band structure of S0/\mose\ and S0/\wse. The 
projection of the energy bands  on S0 [Figs.\,\ref{banda-B_sem}(b2) and (c2)] 
reveals that its metallic bands exhibit a downshift in energy  with respect to 
the Fermi level ($\rm E_F$). This is in agreement with the 
semiconductor\,$\rightarrow$\,S0 net  charge transfers. In addition, we find the 
formation of a metallic band $c1$ for wave vectors parallel to the   $\Gamma$Y 
direction. Further projected energy bands, Figs.\,\ref{banda-B_sem}(b3) and 
(c3), show that the \mose\ and \wse\ MLs also contribute to the formation of 
$c1$. Thus, indicating that the metallic interface states comes from the 
hybridization between the electronic states of the boron sheet S0 and the 
TMDCs. In order to provide further support to the statement above, we have 
performed additional electronic band structure calculations by increasing the 
interlayer distance  $d$ [Fig.\,\ref{hetero}(a2)]. We find that for 
$d$ about 4\,\AA\ the electronic contributions from the \mose\ and \wse\ MLs 
to $c1$  become negligible; it becomes fully occupied and 
localized on the borophene layer. That is, the  metallic interface states in 
S0/TMDC are suppressed by increasing the vertical distance, $d\gtrsim 4$\,\AA.

\begin{figure}
  \includegraphics[width=8.6cm]{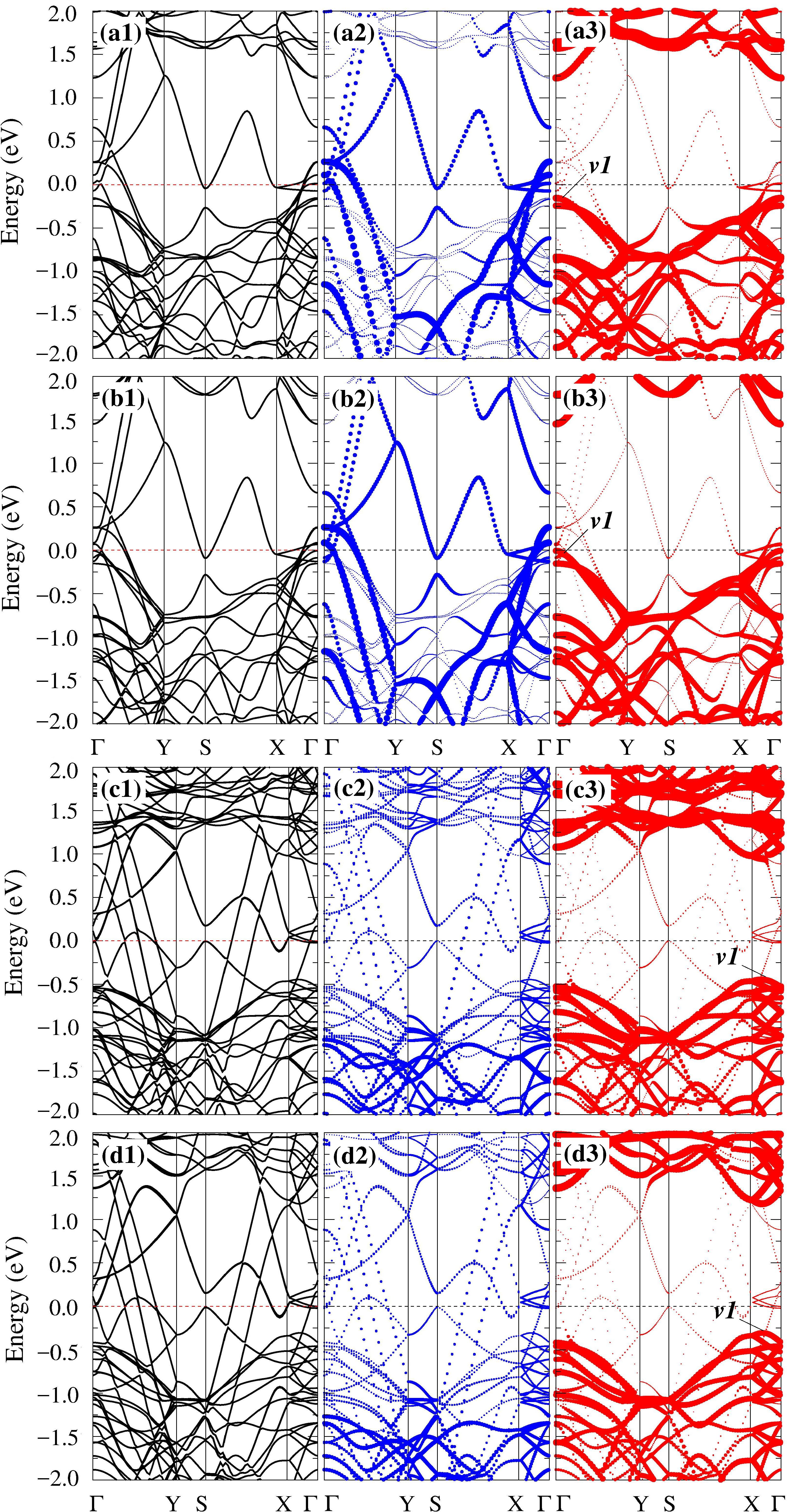}
  \caption{(Color online) Electronic band structures of the S1/\mose\ (a1), 
projected on the S1 sheet (a2), and \mose\ ML (a3); S1/\wse\ (b1), 
projected on the S1 sheet (b2), and \wse\ ML (b3);  S2/\mose\ (c1), 
projected on the S2 sheet (c2), and \mose\ ML (c3); S2/\wse\ (d1), 
projected on the S2 sheet (d2), and \wse\ ML (d3).}
 \label{banda-S1S2_sem}
 \end{figure}
 
Our results of $E^b$  for S1 and S2 lying on the  \mose\ and \wse\ MLs are lower 
when compared with S0/\mose\ and S0/\wse; in addition, the interlayer separation 
increases to $d\approx 3.5$\,\AA, Table~I. Those findings indicate that the 
electronic interaction between the boron sheets  (S1 and S2) and the  
semiconductor \mose\ and \wse\ MLs are weaker compared to the S0/TMDC systems. 
Indeed, this is what we find in Fig.\,\ref{banda-S1S2_sem}, where we can see 
that the energy bands of the separated components are mostly preserved.  Based 
on the Schottky-Mott approach, we find that the \mose\ and \wse\ MLs on S1 
present a p-type Schottky barrier. Here $\Phi_{\rm B}$ is given by the energy 
difference between the  VBM of the semiconductor ML and the Fermi level ($\rm 
E_F$) of the metal/semiconductor-TMDC heterostructure, $\rm \Phi_{\rm B} = E_F 
- E_{\rm VBM}$. For instance, we find a p-type Schottky barrier of  0.15\,eV in 
S1/\mose, while it reduces to $\rm\Phi_B=0.01$\,eV in S1/\wse. It is 
worth noting that, although the structural and energetic similarities with the 
S1/TMDC, the S2/\mose, and S2/\wse\ interfaces present larger values of  
Schottky barrier, {\it viz.}: 0.44 and 0.31\,eV. 

The control of the Schottky barrier  through  an external field, somewhat 
mimicking the gate electrical field,  has been explored in other 
metal/semiconductor 2D systems\,\cite{padilhaPRL2015,liuAPL2016}. Here, based 
on the same approach, we examine the dependence between  $\rm\Phi_B$ and  an 
external electric field perpendicular to the metal/TMDC interface ($E_{\perp}$). 
Here we find that $\Phi_{\rm B}$ increases for $E_{\perp}$ pointing from the 
TMDC ML toward the boron sheet ({\it i.e.} positive values of $E_{\perp}$). 
While for negative values of $E_{\perp}$, the Schottky barrier reduces,  
becoming  negative for $E_{\perp}=-4$\,V/nm in S1/\mose\ and S1/\wse,  giving 
rise to   p-type ohmic contacts. Those results of $\Phi_{\rm B}$ as a function 
of $E_{\perp}$ are summarized in Table~II. It is worth noting that, by 
increasing the strength of $E_{\perp}$, the other S2/\mose\ and S2/\wse\ 
systems will also present p-type ohmic contacts. On the other hand, in contrast 
with the Ref.\,\cite{liuAPL2016},  we did not find any significative change of
$\Phi_{\rm B}$ as a function of $E_{\perp}$; which can be attributed to the 
presence of metallic interface states  pinning  the Fermi level in S0/TMDC.

\begin{table}[h]
\caption{\label{energy} Results of  the Schottky-barrier ($\Phi_{\rm B}$ in 
eV) for S1/\mose, /\wse, S2/\mose, and /\wse, as a function of external 
electric field ($E$ in V/nm).}
\begin{ruledtabular}
\begin{tabular}{lccc}
metal/semic. &  $E=-4$    &  $E=0$ & $E=+4$  \\     
\hline
 S1/\mose\ &  +0.41      &  +0.15   &  $-$0.21   \\
 S1/\wse\  &  +0.32      &  +0.01   &  $-$0.34 \\
 S2/\mose\ &  +0.70      &  +0.44   &  +0.29   \\
 S2/\wse\  &  +0.60      &  +0.31   &  +0.10   \\
\end{tabular}
\end{ruledtabular}
\end{table}

\begin{figure}
  \includegraphics[width=8.6cm]{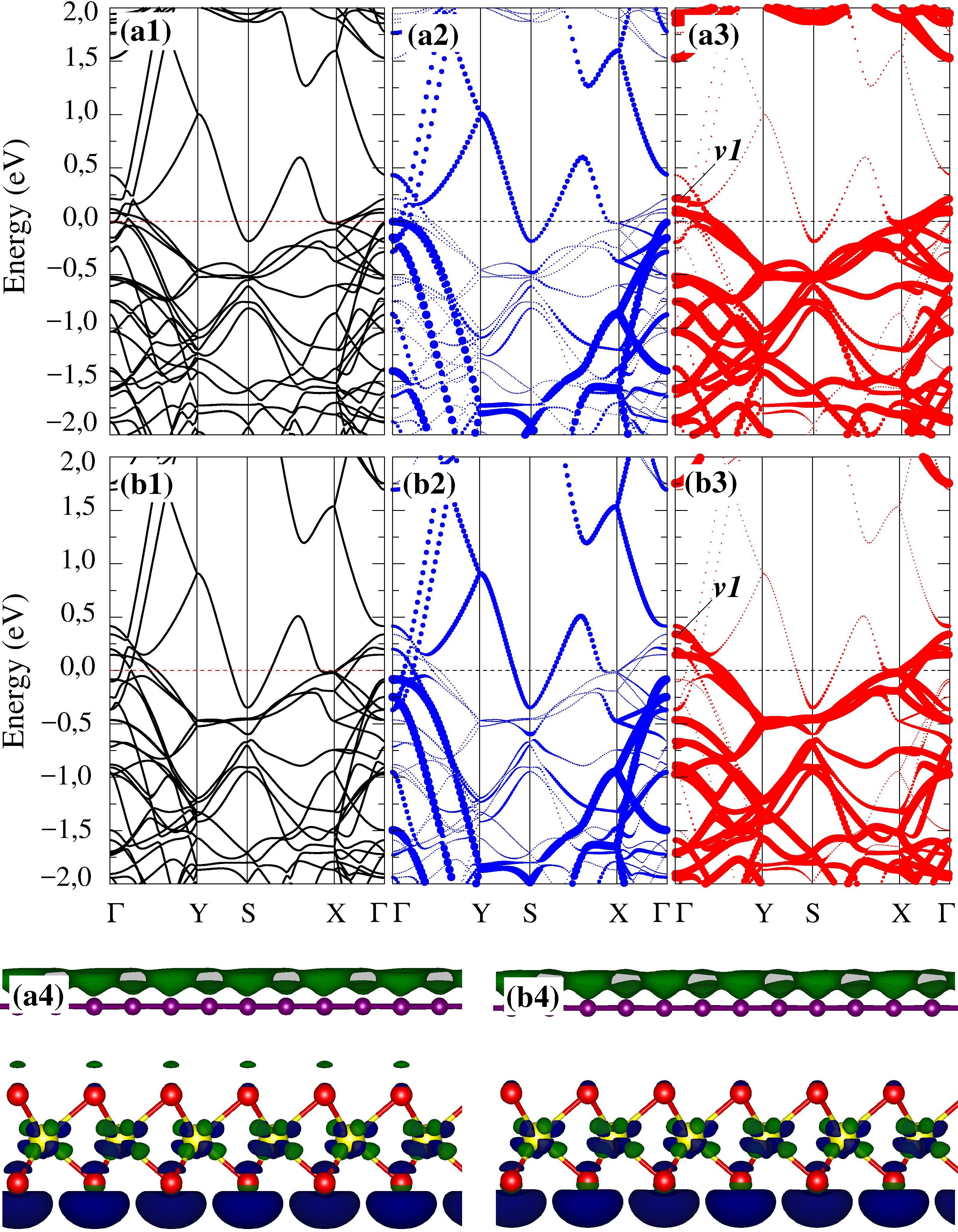}
  \caption{(Color online) Electronic band structure for $E_{\perp}=-4$V/nm, of 
the S1/\mose\ (a1), projected on the S1 sheet (a2), and \mose\ ML (a3); S1/\wse\ 
(b1), projected on the S1 sheet (b2), and \wse\ ML (b3). The total charge 
density redistribution,  $\Delta\rho[E_{\perp}]=\rho[-4\,{\rm V/nm}] - \rho[0]$ 
for S1/\mose\ (a4) and S1/\wse\ (b4). Green regions indicate a net charge gain 
($\Delta\rho>0$), and blue regions indicate a net charge loss ($\Delta\rho<0$), 
with respect to the S1/TMDC system with no external electric field, 
$E_{\perp}=0$. (Isosurfaces of $1.8 \times10^{-4}$ e/\AA$^3$).}
 \label{banda-S1_sem3}
 \end{figure}

The p-type ohmic contact in S1/\mose\  and S1/\wse/  is characterized by net 
charge transfer from the TMDC layer to S1. Indeed, we find a net  charge 
transfers from the \mose, and \wse\ MLs to  S1  of  7.2 and 
$8.5\times10^{12}$e/cm$^2$, respectively. In 
Figs.\,\ref{banda-S1_sem3}(a1)--(a3) and (b1)--(b3) we present the electronic 
band structure of S1/\mose\ and S1/\wse\ upon an external field of $-4$\,V/nm. 
The projected energy bands on the boron sheet S1 [Fig.\,\ref{banda-S1_sem3}(a2)] 
and \mose\ ML [Fig.\,\ref{banda-S1_sem3}(a3)] show that (i) the metallic bands 
of S1 are preserved , however, near the Fermi level, they present a downshift 
when compared with the ones at $E_{\perp}=0$ [Fig.\,\ref{banda-S1S2_sem}(a2)]; 
while (ii) the energy bands projected on the \mose\  is characterized by  the 
formation of partially occupied states near the $\Gamma$-point. Those findings 
[(i) and (ii)] reveal that, indeed,  the S1/\mose\ interface interface exhibits 
a p-type ohmic contact which is tuneable by an external electric field, 
$E_{\perp}$. The same scenario has been verified for the S1/\wse\ interface. In 
Figs.\,\ref{banda-S1_sem3}(a4) and (b4), we present the charge density 
redistribution at the S1/\mose\ and S1/\wse\ interfaces, respectively, as a 
function of the external field, namely $\Delta\rho[E_{\perp}]=\rho[-4\,{\rm 
V/nm}] - \rho[0]$. For both S1/TMDC systems, we find $\Delta\rho[E_{\perp}]<0$ 
mostly localized on the Se layer opposite to the interface region; while the 
boron-sheet S1 presents $\Delta\rho[E_{\perp}]>0$, giving rise to  a net charge 
density depletion at the S1/TMDC interface region.

\section{Conclusions}

Based on the first-principles calculations we have investigated the energetic 
stability and the electronic properties of 2D metal/TMDC vdW heterojunctions. We 
find that the recently synthesized planar boron sheets (S1 and S2) are good 
candidates to form p-type Schottky contacts with the TMDCs (\mose\ and \wse), 
{\it viz.}: S1/\mose,  /\wse, and S2/\mose, and /\wse. There is a net charge 
transfer from the TMDCs to the boron sheets. Upon further application  of 
external electric field, we verify that the S1/\mose\ and /\wse\ heterojunctions 
 exhibit a p-type ohmic contact for $E_{\perp}$ about 4\,V/nm (absolute 
value). 
Similar results are expected for S2/\mose\ and /\wse, however for larger values 
of  $E_{\perp}$. In contrast, such a Schottky barrier tuning has not been 
verified for borophene/TMDCs, S0/\mose\ and S0/\wse. Instead we find the 
formation of metallic 
interface states, pining the Fermi level within the bandgap of the TMDCs.

\section{Acknowledgements}

We would like to thank the Brazilian funding agencies CNPq, CAPES, 
FAPEMIG, and FAPESP. Part of calculations were performed using the 
computational facilities of CENAPAD/SP. 

\section{Appendix}

Since the chemical interactions at the metal/semiconductor interface region are 
negligible, the Schottky-Mott model is quite suitable to estimated the   
Schottky barrier ($\Phi_{\rm B}$) in vdW metal/semiconductor heterojunctions.  
Within such an approach, the p-type Schottky barrier can be written as,
\begin{equation}
 \Phi_{\rm B} = E_{\rm ip} - W,
\end{equation}
where $E_{\rm ip}$  is the ionization potential of the semiconductor, and  $W$ 
represents the metal work function. That is, the Schottky barrier has been 
estimated by comparing  the intrinsic properties of the isolated 
components. 

In order to check the validity of our findings, regarding the formation of 
p-type Schottky barrier in the boron metal/semiconductor heterojunctions studied 
in the present work, we have performed additional electronic structure 
calculations of using the HSE06 hybrid functionals.

\begin{table}[h]
\caption{\label{energy} Ionization potentials ($E_{\rm ip}$ in eV)  and the 
work functions ($W$ in eV) of the pristine systems, TMDCs and boron 
sheets, calculated using the PBE-GGA and HSE06 functionals.}
\begin{ruledtabular}
\begin{tabular}{lccccc}
\multicolumn{1}{c}{}   & 
\multicolumn{2}{c}{$E_{\rm ip}$} &
\multicolumn{3}{c}{$W$} \\
\cline{2-3} \cline{4-6}
 Functional    & \mose\ &  \wse\ & S0  & S1 & S2  \\     
 \hline
 PBE-GGA       & 5.32   & 5.09   &5.28&4.94&4.78 \\
 HSE06         & 5.53   & 5.30   &5.38&4.95&4.79 \\
\end{tabular}
\end{ruledtabular}
\end{table}

Our  PBE-GGA results of the ionization potential, Table~III,  are in good 
agreement with those obtained by Liu {\it et al.}\,\cite{liuSciAdv2016}, while 
$E_{\rm ip}$ calculated using the hybrid functionals (HSE06) increases by about 
0.20\,eV. In this case, we find that, within the HSE06 approach, the  p-type 
Schottky barrier [eq.\,(1)] of S1/\mose\ (S1/\wse) increases from 0.38 to 
0.58\,eV  (0.15 to 0.35\,eV), similarly $\Phi_{\rm B}$ increases by about 
0.2\,eV in  S2/\mose\ and S2/\wse.

\bibliography{/home/hiroki/Trab/RHMiwa}

\end{document}